# HVPE growth of corundum-structured α-Ga$_2$O$_3$ on sapphire substrates with α-Cr$_2$O$_3$ buffer layer

S.I. Stepanov[1,3], V.I. Nikolaev[1,3], A.V. Almaev[2], A.I. Pechnikov[1,3] M.P. Scheglov[1], A.V. Chikiryaka[1], B.O. Kushnarev[2]

[1]Ioffe Institute, 26 Politekhnicheskaya, Saint Petersburg, 194021, Russia

[2]Tomsk state University, 36 Lenin Avenue, Tomsk, 634050, Russia

[3]Perfect Crystals LLC, 28 Politekhnicheskaya Saint Petersburg, 194064, Russia

**Abstract**

Gallium oxide films were grown by HVPE on (0001) sapphire substrates with and without α-Cr$_2$O$_3$ buffer produced by RF magnetron sputtering. Deposition on bare sapphire substrates resulted in a mixture of α-Ga$_2$O$_3$ and ε-Ga$_2$O$_3$ phases with a dislocation density of about $2 \cdot 10^{10}$ cm$^{-2}$. The insertion of α-Cr$_2$O$_3$ buffer layers resulted in phase-pure α-Ga$_2$O$_3$ films and a fourfold reduction of the dislocation density to $5 \cdot 10^9$ cm$^{-2}$.

**Introduction**

Gallium oxide (Ga$_2$O$_3$) has attracted considerable attention owing to its ultra-wide bandgap of about 4.9 eV and a high breakdown electric field of 8 MV/cm. Ga$_2$O$_3$ is known to exist in at least five different crystal structures, specifically, α-, β-, δ-, ε (κ)-, and γ-phases. The metastable α-Ga$_2$O$_3$ polymorph is of particular interest because of its wider bandgap (~5.3 eV) and corundum-type crystal structure with lattice parameters a=b=4.9825Å and c=13.433 Å [1]. Thin films of α-Ga$_2$O$_3$ can be synthesized at relatively low temperatures by epitaxial growth on single-crystal substrates.

Epitaxial films of α-Ga$_2$O$_3$ are most commonly grown on sapphire substrates because they are readily available, relatively inexpensive, and share the same corundum structure as α-Ga$_2$O$_3$. However, the lattice mismatch of α-Ga$_2$O$_3$ and α-Al$_2$O$_3$ is as high as 4.8%, which results in a high density of threading dislocations in α-Ga$_2$O$_3$ layers. Although the HVPE growth of phase-pure α-Ga$_2$O$_3$ films on sapphire has been achieved by a number of researchers [2,3], phase control in Ga$_2$O$_3$ films remains a rather challenging task. The growth conditions α-Ga$_2$O$_3$ and ε-Ga$_2$O$_3$ are very similar, resulting in concurrent growth of α- and ε- phases [4].

Common approaches to reduce the dislocation density in heteroepitaxial Ga$_2$O$_3$ films are epitaxial lateral overgrowth [5], growth on patterned substrates [6,7], and insertion of a buffer layer. So far, epitaxial growth of Ga$_2$O$_3$ on sapphire using α-(AlGa)$_2$O$_3$ [8,9], α-Fe$_2$O$_3$ [10] buffer layers has been reported. Chromium oxide (α-Cr$_2$O$_3$) is another suitable buffer layer material. As pointed out by Kaneko *et al.* [11], α-Cr$_2$O$_3$ has the same corundum-type structure with lattice parameters $a=b=4.9607$ Å and $c=13.599$ Å [12]. The lattice mismatch between α-Ga$_2$O$_3$ and α-Cr$_2$O$_3$ along the *a*-axis is only 0.4% [11].

In this paper, we compare Ga$_2$O$_3$ films deposited in the same growth runs on sapphire substrates with and without the Cr$_2$O$_3$ buffer layer.

**Experiment**

Thin Cr$_2$O$_3$ films used in this study were deposited in an RF magnetron sputtering system (A-500 Edwards, UK). C-plane sapphire wafers of 330 - 440 μm in thickness and 50 mm in diameter were used as substrates. Before deposition, the sapphire substrates were cleaned in sulfuric acid and isopropanol. A 99.95% chromium target and oxygen-argon plasma were used as source materials. The concentration of oxygen in the plasma was 56.1 ± 0.5 vol.%. The distance between the substrate and the target was 70 mm. The deposition was carried out at the working pressure of $7 \cdot 10^{-3}$ mbar and the RF power of 70 W. The temperature of the substrate was not controlled during the sputtering. The deposition was carried out for 45 minutes to reach a thickness of about 150 nm. After deposition, the Cr$_2$O$_3$ films were annealed in air at 500 °C for 3 hours.

Ga$_2$O$_3$ films were grown in a homemade atmospheric horizontal quartz HVPE reactor. Gallium chloride (GaCl) and oxygen (O$_2$) were used as precursors. GaCl was synthesized *in situ* by passing gaseous hydrogen chloride (HCl, 99.999% pure) over metallic gallium (Ga, 99.9999% pure) at 600 °C. The GaCl and $O_2$ were then mixed in the deposition zone of the reactor to produce $Ga_2O_3$ on the substrate. Argon was used as a carrier gas to keep the total gas flow rate through the reactor at 10 slm. The deposition temperature was 500 °C. The VI/III ($2O_2$/GaCl) ratio was 4.2. Under these conditions, the growth rate was 2.4 µm/h. To ensure identical growth conditions and fair comparison, the growth on sapphire substrates with and without $Cr_2O_3$ buffer was conducted in the same growth run.

The phase composition of the produced films was investigated by x-ray diffraction (XRD) analysis using CuK$\alpha_1$ radiation. The structural quality of the films was estimated from the full width at half maximum (FWHM) of XRD rocking curves (RC) of symmetric (0006) and skew-symmetric ($10\bar{1}8$) reflections of α-$Ga_2O_3$.

**Results and discussion**

Even though the chromium oxide films were deposited by the RF magnetron sputtering at low temperature, they exhibited, rather surprisingly, a high degree of crystallinity. Only the diffraction peaks from α-$Cr_2O_3$ and α-$Al_2O_3$ (0001) planes are present in the XRD 2θ–ω scan (Figure 1a) which indicates that the films are (0001) oriented α-$Cr_2O_3$ with corundum structure.

Figure 1b shows a representative XRD ω-2θ scan for the $Ga_2O_3$ layer deposited by HVPE on a bare sapphire. As one can see, both α- and ε-phases are present in the film. The results are in line with earlier studies where the HVPE growth of α-$Ga_2O_3$ on *c*-plane sapphire was hindered by the formation of the ε-phase [13]. The in-plane lattice mismatch between ε-$Ga_2O_3$ and sapphire is 4.1% [8] while α-$Ga_2O_3$ has the lattice mismatch of 4.8% [14]. Therefore it can be speculated that the ε-phase on c-plane sapphire is energetically more favorable than the α-phase.

In contrast, $Ga_2O_3$ layer deposited in the same growth run on the sapphire substrate with α-$Cr_2O_3$ buffer was found to be a pure α-phase as only diffractions from α-$Ga_2O_3$, α-$Cr_2O_3$, and α-$Al_2O_3$ can be seen in the XRD ω-2θ scan (Figure 1c).

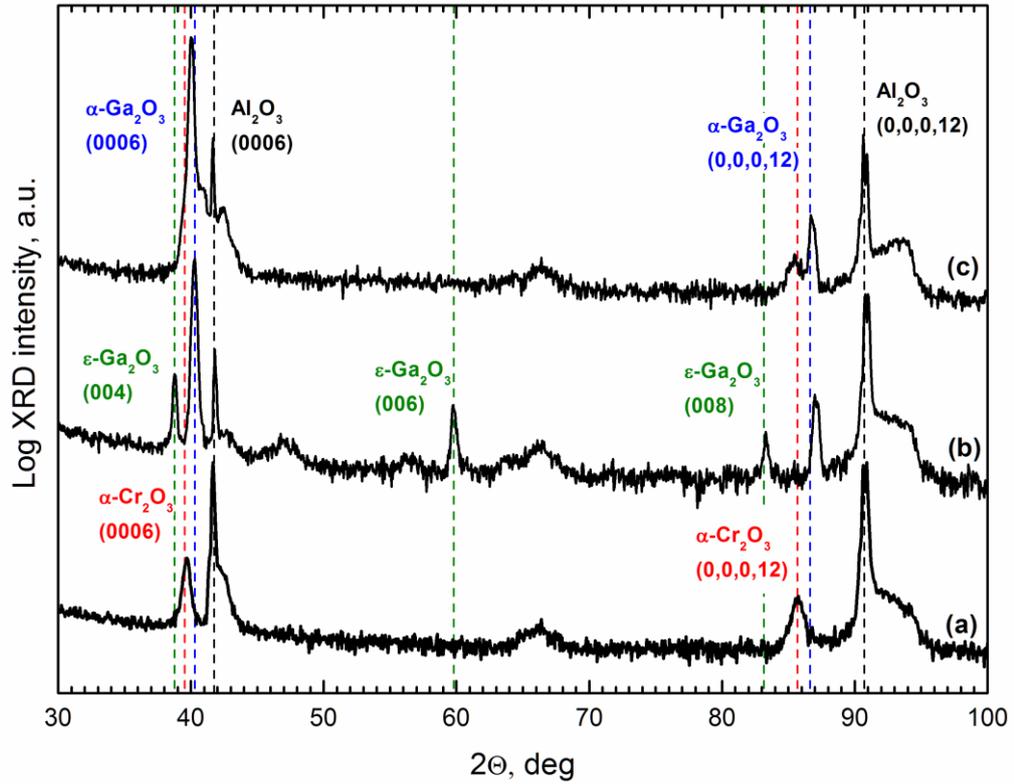

**Figure 1.** XRD 2θ/ω scan profiles of (a) α-Cr$_2$O$_3$ film deposited on sapphire, (b) HVPE Ga$_2$O$_3$ on sapphire, (c) HVPE Ga$_2$O$_3$ on sapphire substrate with the α-Cr$_2$O$_3$ buffer layer.

It is noteworthy that α-Ga$_2$O$_3$ films on bare sapphire substrates exhibited narrow RC for the (0006) reflection with FWHM of about 60 arcsec and much broader RC for the ($10\bar{1}8$) reflection with FWHM of about 2400 arcsec. Such dramatic difference between the FWHM values for symmetric and asymmetric rocking curves is typical for α-Ga$_2$O$_3$ films on sapphire and has been reported by other authors for α-Ga$_2$O$_3$ films prepared by HVPE [15], mist-CVD [16], and MBE [17] techniques. In contrast, α-Ga$_2$O$_3$ films grown on sapphire with α-Cr$_2$O$_3$ buffer showed very similar FWHM values for (0006) and ($10\bar{1}8$) diffractions of about 1400 arcsec each.

The FWHM of symmetric scans is indicative of the tilt mosaic, while skew symmetric RC are broadened both by tilt and twist mosaics. Assuming that the tilt and twist components are independent and follow normal distributions, the FWHM of a skew symmetric reflection $\beta_{skew}$ can be expressed as a convolution of tilt and twist components:

$$\beta_{skew}^2 = (\beta_{tilt} \cos \varphi)^2 + (\beta_{twist} \sin \varphi)^2 \quad (1)$$

where $\beta_{tilt}$ and $\beta_{twist}$ are the FWHMs of the tilt and twist angle distributions, and $\varphi$ is the angle between the inclined plane and the sample surface. The tilt angle can be approximately taken as the FWHM of the symmetric (0006) reflection $\beta_{tilt} \approx \beta_{(0006)}$. Then the twist angle can be calculated from the FWHM of asymmetric RC by using the formula 1.

The tilt and twist mosaics are associated with the densities of the screw ($D_s$) and edge ($D_e$) dislocations, respectively. The dislocations densities can be roughly estimated from the following relationships:

$$D_e = \frac{\beta_{twist}^2}{4.35 b_e^2} \qquad D_s = \frac{\beta_{tilt}^2}{4.35 b_s^2} \quad (2)$$

where $b_s = \langle 0001 \rangle = 13.43$ Å and $b_e = \frac{1}{3}\langle 2\bar{1}\bar{1}0 \rangle = 4.98$ Å are the lengths of the Burgers vectors for screw and edge dislocations, respectively. The dislocation densities in the α-Ga$_2$O$_3$ films grown on bare sapphire can be estimated as $D_s = 1 \cdot 10^6$ cm$^{-2}$ and $D_e = 2 \cdot 10^{10}$ cm$^{-2}$ for screw and edge dislocations, respectively. The total density of threading dislocations $D_{total}$ is $2 \cdot 10^{10}$ cm$^{-2}$. The films grown on α-Cr$_2$O$_3$ buffer have a much higher screw dislocation density of $D_s = 6 \cdot 10^8$ cm$^{-2}$. However, the density of the edge dislocations was reduced to $D_e = 4 \cdot 10^9$ cm$^{-2}$ resulting in a more than fourfold decrease in the total dislocation density is $D_{total} \sim 5 \cdot 10^9$ cm$^{-2}$.

**Conclusions**

The HVPE growth of α-Ga$_2$O$_3$ on sapphire substrates with and without α-Cr$_2$O$_3$ buffer layer has been investigated. The α-Cr$_2$O$_3$ buffer was found to be an effective method to produce phase pure α-Ga$_2$O$_3$ films. It was found that Ga$_2$O$_3$ films grown on bare sapphire substrates had a low density of the dislocations with a screw component and a high density of pure edge dislocations. In contrast, α-Ga$_2$O$_3$ films grown on the α-Cr$_2$O$_3$ buffer exhibited a more balanced ratio of screw and edge dislocations. The insertion of the α-Cr$_2$O$_3$ buffer layer reduced the threading dislocation density from 2·10$^{10}$ cm$^{-2}$ to 5·10$^9$ cm$^{-2}$. It is believed that the quality of the α-Ga$_2$O$_3$ films can be improved further by optimizing the growth parameters of the α-Cr$_2$O$_3$ buffer.

**Acknowledgments**

The study was supported by Perfect Crystals LLC.